\documentstyle[12pt,epsf]{article}
\renewcommand{\thefootnote}{\alph{footnote}}

\begin{document}

\begin{tabbing}
\` Nevis R\#1502
\end{tabbing}

\begin{center}
\bf Determination of the Strange Quark Content of the Nucleon from \\
 a Next-to-Leading-Order QCD Analysis of Neutrino Charm Production        
\end{center}
\vskip\baselineskip

\vskip .2in

\parskip 0.12in
\centerline{
A.O. Bazarko, C.G. Arroyo, 
K.T. Bachmann,\footnote{Present address: National Center for Atmospheric
Research,
Boulder, CO 80307.}
 T. Bolton, 
C. Foudas,\footnote{Present address: University of Wisconsin, Madison, WI
53706.} 
B.J. King,\footnote{Present address: CERN, CH-1211 Geneva 23, Switzerland.}
}
\centerline{
 W.C. Lefmann, W.C. Leung,
S.R. Mishra,\footnote{Present address: Harvard University,
Cambridge, MA 02138.}
E. Oltman,\footnote{Present address: Lawrence Berkeley
Laboratory, Berkeley, CA 94720.}
P.Z. Quintas,\footnote{Present address: Fermilab, Batavia, IL
60510.}
}
\centerline{ S.A. Rabinowitz, 
F.J. Sciulli, W.G. Seligman, M.H. Shaevitz
}
\centerline{\bf Columbia University, New York, NY 10027}

\centerline{ 
F.S. Merritt, M.J. Oreglia, 
B.A. Schumm,$^e$ 
}
\centerline{\bf University of Chicago, Chicago, IL 60637}

\centerline{ R.H. Bernstein, F. Borcherding, H.E. Fisk, M.J. Lamm,}
\centerline{
W. Marsh, K.W.B. Merritt, 
H.M. Schellman,\footnote{Present address: Northwestern University, 
Evanston, IL 60208.}
D.D. Yovanovitch
} 
\centerline{\bf Fermilab, Batavia, IL 60510}

\centerline{
A. Bodek, H.S. Budd, 
P. de Barbaro, W.K. Sakumoto
}
\centerline{\bf University of Rochester, Rochester, NY 14627}

\centerline{
T. Kinnel, P.H. Sandler,\footnote{Present address: Lawrence
Livermore National Laboratory, Livermore, CA  94550.}
W.H. Smith
}
\centerline{\bf University of Wisconsin, Madison, WI 53706.}

\vskip\baselineskip
\centerline{\bf (CCFR Collaboration)}

\vskip1in
\begin{center}
 \today. \\
Submitted to Physics Letters B.
\end{center}
\newpage
\renewcommand{\thefootnote}{\arabic{footnote}}
\setcounter{footnote}{0}

{\hspace*{0.3cm}

We present the first next-to-leading-order QCD analysis of neutrino charm 
production, using a sample of 6090 $\nu_\mu$- and $\overline\nu_\mu$-induced
opposite-sign  dimuon events observed in the CCFR detector at the Fermilab
Tevatron. We find that the nucleon strange quark content is suppressed  with
respect to the non-strange sea quarks by a factor $\kappa = 0.477 \:
^{+\:0.063}_{-\:0.053}$, where the error includes statistical, systematic and
QCD scale uncertainties. In contrast to previous leading order analyses, we
find that the strange sea $x$-dependence is similar to that of the non-strange
sea, and that the measured charm quark mass, $m_c = 1.70 \pm 0.19 \:{\rm
GeV/c}^2$,  is larger and consistent with that determined in other processes. 
Further analysis finds that  the difference in $x$-distributions between
$xs(x)$ and $x\overline s(x)$ is small. A measurement of the
Cabibbo-Kobayashi-Maskawa matrix element $|V_{cd}|=0.232
\;^{+\:0.018}_{-\:0.020}$ is also presented. 

}

\vskip .1in

\vfill\eject

\noindent{\bf 1. Introduction}

Nucleon structure at high momentum transfers 
is characterized by parton distribution functions, which 
describe the proton and neutron in terms of quarks and
gluons using the factorization 
theorems of Quantum Chromodynamics (QCD) \cite{PQCD}. 
These nucleon parton distributions are essential inputs when using
perturbative calculations to predict high energy processes involving nucleons,
such as those at the Tevatron collider and the planned LHC.
The ability to consistently predict such processes using 
one set of universal parton distributions
is an important test of QCD as the theory of the strong interactions.
Neutrino-nucleon deep-inelastic scattering is particularly suited for 
measuring the parton densities due to 
the neutrino's ability to resolve the flavor of the nucleon constituents. 
Furthermore, neutrino scattering is an effective way 
to study the dynamics of heavy quark production, due to the light to heavy
quark transition at the charged current vertex.
In particular, neutrino charm
production can be used to isolate the nucleon
strange quark distributions, $xs(x)$ and $x\overline s(x)$. 

We present the first next-to-leading-order (NLO) QCD analysis of
neutrino and antineutrino production of charmed quarks 
including the
first direct determination of the strange quark distribution defined at NLO.
The order of the strange quark distribution from this analysis
matches that of recent global nucleon structure analyses, 
like those of CTEQ \cite{cteq} and MRS \cite{mrs}.
Since the data presented here provide the most sensitive constraints on 
the strange quark
distribution, these results should become important ingredients in 
future global parton distribution fits.

In addition, we present measurements of
the Cabibbo-Kobayashi-Maskawa (CKM) 
matrix element $|V_{cd}|$ and the mass of the charm quark $m_c$.   
The charm quark mass from this analysis 
is directly comparable with $m_c$ measurements derived from NLO
analyses of other processes. Such comparisons are a test 
of the perturbative QCD phenomenology for heavy quarks. 
Neutrino production of charm off nucleon $d$ quarks 
provides a direct determination of the product $|V_{cd}|^2 B_c$, 
where $B_c$ is the weighted average
of the semi-leptonic branching ratios of the charmed hadrons produced. 
$|V_{cd}|$ is isolated by including an estimate of $B_c$ 
using data from other experiments, including a re-analysis of Fermilab 
neutrino-emulsion data.

The signature
for the production of charmed quarks in neutrino- and antineutrino-nucleon
scattering is the presence of two oppositely-signed muons.  
In the case of neutrino scattering, the underlying process is a neutrino
interacting with an $s$ or $d$ quark, producing a charm quark that
fragments into a charmed hadron. The charmed hadron's semileptonic decay
produces a second muon of opposite sign from the first. 
\begin{eqnarray}
\nu_\mu\; + \; {\rm N} \;
 \longrightarrow \; \mu ^{-} \!\! & + & \! c \; +\; {\rm X}
\nonumber \\
& & \!\!\hookrightarrow \;  \mu ^{+} \; + \; \nu_\mu \nonumber
\end{eqnarray}
The analogous process
with an incident antineutrino proceeds through an interaction with 
an $\overline s$
or $\overline d$ antiquark, again leading to oppositely-signed 
muons in the final state.
\begin{eqnarray}
\overline\nu_\mu\; + \; {\rm N} \;
 \longrightarrow \; \mu ^{+} \!\! & + & \! \overline c \; +\; {\rm X}
\nonumber \\
& & \!\! \hookrightarrow \;  \mu ^{-} \; + \; \overline\nu_\mu \nonumber
\end{eqnarray}

\noindent{\bf 2. The CCFR detector and event selection}

The dimuon data were accumulated 
during two runs, E744 and E770, 
with the Chicago-Columbia-Fermilab-Rochester (CCFR) detector
at the Fermilab Tevatron Quad-Triplet neutrino beam.
This wide-band beam was composed of $\nu_\mu$ and
$\overline\nu_\mu$ with energies up to 600 GeV and a flatter energy spectrum 
than characteristic of horn-focused neutrino beams.
In the CCFR detector \cite{sakumoto,king},
neutrino interactions occur in the 690 ton unmagnetized steel-scintillator
target-calorimeter, which is instrumented with drift chambers for muon 
tracking. The calorimeter's hadronic energy resolution is
$\sigma/E = 0.85/\sqrt{E}$. The target is followed by a solid-iron 
toroidal magnetic spectrometer, which
identifies muons and measures their momenta with a 
resolution $\Delta p = 0.11 p$. 

The detector measures 
$p_{\mu_1}$ and $p_{\mu_2}$, the momenta of the two muons, 
$\theta_{\mu_1}$ and $\theta_{\mu_2}$, the
angles of the muons with respect to the neutrino beam axis, 
and $E_{\rm had}^{\rm vis}$, the
energy of the hadronic system. 
Muon 1 is labeled as the leading muon, defined as 
the one emerging from the leptonic vertex, using a procedure described below.
The visible  
energy in a dimuon event, 
$E_{\rm vis}=E_{\mu_1}+E_{\mu_2}+E_{\rm had}^{\rm vis}$,
misses the energy of the decay neutrino. 
Therefore, we
make the distinction between the {\em visible} quantities---quantities derived
directly from measurement---and the {\em physical} quantities, 
which are inferred on average by correcting distributions of the visible
quantities using the Monte Carlo simulation described below. 
Variables commonly used to describe deep
inelastic scattering are: $Q^2_{\rm vis}=4E_{\rm vis}E_{\mu_1}{\rm
sin}^2(\theta_{\mu_1}/2)$, the negative square of the four-momentum transfer, 
$x_{\rm vis}=Q^2_{\rm vis}/[2M(E_{\rm had}^{\rm vis}+E_{\mu_2})]$, 
the Bjorken scaling
variable, where $M$ is the nucleon mass, $y_{\rm vis}=(E_{\rm had}
+E_{\mu_2})/E_{\rm vis}$, the inelasticity, and 
$W^2_{\rm vis}=M^2+Q^2_{\rm
vis}(1/x_{\rm vis}-1)$, the invariant mass squared of the hadronic system.

In the experiment, charged-current single muon events
are required to have 
$E_{\rm vis}>30$ GeV, $E_{\rm had}^{\rm vis}>10$ GeV,
$Q^2_{\rm vis}>1$ GeV$^2$/c$^2$ and 
$p_{\mu_1}>9$ GeV/c.
Dimuon events are selected by making the further requirement 
that the second muon has $p_{\mu_2}>5$ GeV/c and that both muons have 
$\theta_\mu<0.250$ rad.
The second muon's momentum is measured in the magnetic
spectrometer whenever possible, otherwise 
it is determined from the muon's range in the target. 
In order to reduce non-prompt sources of second muons, events in which 
muon 2 does not reach the toroid
must also satisfy $E_{\rm had}^{\rm vis}<130$ GeV.
The final dimuon sample contains 6090 events and 
is characterized by 
$\langle E_{\rm vis}\rangle =192$ GeV, 
$\langle W^2_{\rm vis}\rangle =168$ ${\rm GeV^2/c^2}$, 
$\langle Q^2_{\rm vis} \rangle = 25.5$ ${\rm GeV^2/c^2}$,
and $\langle x_{\rm vis}\rangle =0.15$. 
Results from a leading-order analysis of this data
sample were reported previously \cite{sar}.

\pagebreak
\noindent{\bf 3. Differential cross section}

The differential cross section for dimuon
production is expressed generally as
\begin{equation}
{d^3\sigma (\nu_\mu N\rightarrow \mu^-\mu^+ X) \over d\xi\: dy\: dz } =
{d^2\sigma (\nu_\mu N\rightarrow cX) \over d\xi\: dy} 
\: D(z) \: B_c(c\rightarrow \mu^+ X), 
\label{eq:dimuon}
\end{equation}
where the function $D(z)$ describes the hadronization of charmed quarks and 
$B_c$ is the weighted average of the semi-leptonic branching ratios 
of the charmed hadrons produced in neutrino interactions.

The heavy charm quark introduces an energy threshold in the
charm production rate. This is a kinematic effect for which  $\xi$, 
the momentum fraction of the struck quark, is related to the Bjorken 
scaling variable $x$, through the expression \cite{AOT2}
\begin{equation}
\xi = \left({1\over 2x} + \sqrt{{1\over 4x^2}+{M^2\over Q^2}}\right)^{-1} 
{ Q^2-m_s^2+m_c^2+\Delta \over 2Q^2}  ,
\end{equation}
where $m_c$ is the charm quark mass and $m_s$ refers to the initial state quark 
mass, either the strange quark or the down quark, 
and
$\Delta=\Delta(-Q^2,m_s^2,m_c^2)$ is the triangle function, defined by
$\Delta(a,b,c)\equiv\sqrt{a^2+b^2+c^2-2(ab+bc+ca)}$. 
The full expression for $\xi$ can be simplified  
by neglecting the small effect of the initial state quark mass to yield
\begin{equation}
\xi = x\left(1 + {m_c^2\over Q^2}\right) 
\left(1 - {x^2M^2\over Q^2}\right).
\end{equation}
Relating $\xi$ and $x$ through the charm quark mass 
is referred to as slow-rescaling \cite{barnett}.

At leading order (LO) charm is produced by scattering directly 
off of strange and down quarks in the nucleon. The 
LO differential cross section 
for an isoscalar target, neglecting target mass effects, is given by:
\begin{eqnarray}
\left\{ {d^2\sigma (\nu_\mu N\rightarrow cX) \over d\xi\: dy}\right\}_{LO} 
& = & 
\frac{G^2ME_\nu }{\ \pi (1+Q^2/M_W^2)^2 } \; \{ \; [\xi u(\xi,\mu^2)+  
\xi d(\xi,\mu^2 )]\: |V_{cd}|^2  \nonumber \\  
 &  & + \; 2\xi s(\xi,\mu^2) \:
|V_{cs}|^2 \; \} \left( 1-\frac{m_c^2}{\ 2ME_\nu \xi }\right), 
\label{eq:lo}
\end{eqnarray}
where $\xi u(\xi,\mu^2 )$, $\xi d(\xi,\mu^2)$ and $\xi s(\xi,\mu^2)$ represent 
the momentum distributions of the $u$, $d$ and $s$ quarks within 
the proton 
(the corresponding $\overline{\nu }_\mu $ process has the 
quarks replaced by their antiquark partners)
and $|V_{cd}|$ and $|V_{cs}|$ are the CKM matrix elements.
The dependence of the parton distributions on the scale $\mu^2$ is
specified by QCD \cite{glap}.
In the leading order analysis of Reference \cite{sar}, 
Callan-Gross violation is included by replacing the term
$\left[1-m_c^2/(2ME_\nu\xi)\right]$ in Equation (\ref{eq:lo}) with 
$\left\{[1+R_L(\xi,\mu^2)]\right.$
$[1+(2M\xi /Q)^2]^{-1}$ $[1-y-Mxy/(2E)]$ $\left.+xy/\xi \right\}$, 
and using external measurements   
of the structure function $R_L(\xi,\mu^2)$ \cite{rl}. 
In the NLO formalism, violation of the Callan-Gross relation  
emerges as a consequence of QCD.
 
The LO expression illustrates the sensitivity of the process to the 
strange quark sea. Charm (anticharm) production from scattering off $d$
$(\overline d)$ quarks is Cabibbo
suppressed. In the case of charm produced by neutrinos, 
approximately 50\% 
is due to scattering from $s$ quarks, even though the $d$ quark content of
the proton is approximately ten times larger. In the case of
antineutrino scattering, where $\overline d$ quarks from the sea contribute,
roughly 90\% is due to scattering off $\overline s$ quarks. 

\begin{figure}[h,b,t]
\begin{minipage}[t]{78mm}
   \epsfysize=6cm \epsfbox[82 240 508 576]{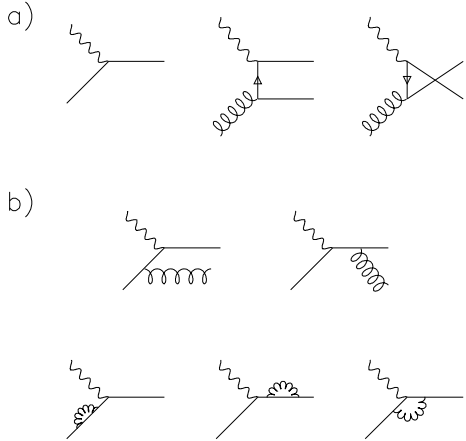}
   \caption[]{
   Mechanisms that contribute to neutrino production of charm up to  
${\cal O}(\alpha_S)$. {\bf a)} The dominant diagrams: the leading-order
quark-initiated diagram, and the t channel and u channel gluon-initiated
diagrams, respectively. {\bf b)} The radiative-gluon and self-energy diagrams.
   }
\label{fig:NLO}
\end{minipage}
\end{figure}

Because neutrino charm production has a large sea quark  
component at leading-order, 
the next-to-leading-order 
gluon-initiated contributions  
are significant \cite{AOT}. The size of the gluon distribution, 
which is an order of
magnitude larger than the sea quark distribution, compensates for the extra
power of $\alpha_S$ involved in the gluon-initiated diagram.
The NLO quark-initiated diagrams, shown in Figure \ref{fig:NLO}b, 
in which a gluon is radiated, also enter the perturbative expansion at ${\cal
O}(\alpha_S)$, but the contributions of these diagrams 
to the cross section are not enhanced by large underlying 
parton distributions.
Calculations including the next-to-leading-order formalism
have recently become available \cite{collins,kramer,gj} and lead 
to the analysis in this Letter.

\pagebreak
\noindent{\bf 4. Monte Carlo simulation}

Information about the strange sea, the charm quark 
mass, and the branching ratio is extracted by comparing 
the $x_{\rm vis}$ and $E_{\rm vis}$ distributions of 
the data to theoretical expectations contained in a Monte Carlo simulation.
The Monte Carlo program models the dependence of these physics parameters 
as well as the effects 
of detector acceptance, resolution smearing, and
missing energy associated with the charmed particle decay.
Dimuon Monte Carlo event generation proceeds by 
using a simulated sample of charged-current single muon events 
and demanding that the hadronic system
contains a second muon from charm decay.
The single muon Monte Carlo sample is  
normalized to the charged-current data sample, ensuring that 
the Monte Carlo energy spectrum exactly models that of the data.

The species of charmed particles produced in neutrino
interactions as a function of neutrino energy was measured by Fermilab 
E531 \cite{e531}. 
With an $E_\nu>30$ GeV cut, the
production is dominated by charged and neutral $D$ mesons.
Fragmentation to $D$'s in the Monte Carlo simulation is parameterized by 
the fragmentation function of Collins and Spiller \cite{colfrag},
 $D(z) = N \left[(1-z)/z + \epsilon(2-z)/(1-z)\right] (1+z)^2 
[1-(1/z)-\epsilon /(1-z)]^{-2}$,  
where $z=p_D/p^{\rm max}_D$ is the fraction of its maximum momentum that the 
$D$ meson carries and $\epsilon$ is a free parameter. 
The parameter $\epsilon$ in the fragmentation model is fit by using the 
distribution of $z_{\rm vis}=E_{\mu_2}/(E_{\mu_2}+E_{had})$.  

The dimuon events are divided into those from incident $\nu_\mu$ or 
$\bar\nu_\mu$ by a separation procedure
that assumes that the leading muon has larger transverse momentum
with respect to the direction of the hadron shower than the muon from the 
charmed
hadron decay. This procedure separates the sample into 5030 $\nu_\mu$-induced
events and 1060 $\overline\nu_\mu$-induced events.
The largest uncertainty in this procedure is due to the knowledge of the charm
meson $p_\perp$ distribution. For this study we parameterize the $p_\perp$
distribution by 
$d{\rm n}/dp_\perp^2 \propto e^{-\beta p_\perp^2}$ and 
use the Fermilab E531 emulsion data \cite{e531} with
$W^2>30$ GeV$^2$/c$^2$ to determine  $\beta = 1.21\pm 0.34$.
Using this method, the separation procedure is found to misidentify 
$5.8 \pm 0.4$\% of the $\nu$ events and 
$7.3 \pm 0.4$\% of the $\overline\nu$ events. 

The charm-initiated dimuon signal is contaminated by non-prompt pion and kaon
decay. The high-density calorimeter 
minimizes this contamination due to the short interaction length of the
detector. A combination of hadronic test beam muoproduction data
and Monte Carlo simulations predicts a small $\pi$/K decay background of 
$797\pm 118$ $\nu_\mu$ and $118\pm 25$ $\overline\nu_\mu$ 
events \cite{pamela}.

To calculate the probability of producing charm, we employ
the NLO QCD charm production differential cross section calculation 
of Aivazis, Collins, Olness and Tung \cite{collins},  
including the Born and gluon-fusion diagrams, shown in Figure \ref{fig:NLO}a. 
These are the leading contributions to charm production.
The calculation is performed
in the $\overline{\rm MS}$ scheme. The factorization scale in the calculation 
is chosen to be $\mu=2p_\perp^{\rm max}$,
where $p_\perp^{\rm max}=\Delta(W^2,m_c^2,M^2)/\sqrt{4W^2}$ 
is the maximum available transverse momentum of the initial state quark coming 
from the gluon splitting, or
equivalently of the final state charm quark, for the given kinematic variables 
$x$ and $Q^2$. The renormalization scale is chosen to equal the factorization
scale. We discuss the uncertainty due to the choice of these
scales below. Electromegnetic radiative corrections to the cross section 
are calculated using the method of Bardin {\em et al.} \cite{bardin}.

The finite momentum cut on the second muon limits the acceptance of events
attributable to gluon-initiated production, 
$$
W^+ g \rightarrow c \overline s. 
$$
Gluon-initiated production of charm proceeds through both the t and u 
channels as shown in Figure \ref{fig:NLO}a.  
While these diagrams are quantum mechanically equivalent, they dominate
different regions of phase space. 
In the t channel, the gluon splits into an $s \overline s$
pair and the $c$ quark emerges from the W-boson vertex.
In the u channel,
the legs of the $c$ and $\overline s$ quarks 
are crossed---the gluon splits into a $c\overline c$ pair and the
$\overline s$ quark emerges from the W-boson vertex.

In the W-boson--gluon center of mass frame, 
the $c$ quark is produced at an angle $\theta^\ast_c$
relative to the W-boson direction. The production angle is related to the
momentum of the $c$ quark in the lab, and hence with $p_{\mu_2}$. 
When $\theta^\ast_c$ is small---t channel dominance---the 
$c$ quark carries most of the W-boson momentum. 
As $\theta^\ast_c$ approaches $\pi$---u channel dominance---the 
$c$ quark emerges with little momentum in the lab. 
Consequently, events with large $\theta^\ast_c$ 
are less likely to produce a second 
muon with $p_{\mu_2} > 5$ GeV/c.

An acceptance correction due to this effect 
is determined by folding the calculated squared production amplitude
for producing charm at angle $\theta^\ast_c$ with the experimental acceptance.
To determine the experimental acceptance,
the ability of events with finite $\theta^\ast_c$ to pass the 5 GeV/c 
$p_{\mu_2}$ cut is
compared to that when the events are generated with $\theta^\ast_c=0$.
We find that this relative acceptance drops to near zero at 
$\theta^\ast_c=\pi$, and is about 74\% at $\theta^\ast_c=\pi/2$. 
Integrating over the calculated squared production amplitude, which peaks in
the forward and backward directions, 
the overall t--u channel acceptance correction for
gluon-initiated production is $60\pm 10$\%.
The effect of this acceptance correction is small but not 
insignificant; for example, it shifts the value of $m_c$ 
determined from the fit described below by $+0.07$ GeV/c$^2$.

Measurements of the $F_2$ and $xF_3$ structure functions 
by CCFR \cite{quintas,leung} are used
to determine the singlet and the non-singlet quark 
distributions,
$xq_{SI}(x,\mu^2)=xq(x,\mu^2)+x\overline q(x,\mu^2)$ and 
$xq_{NS}(x,\mu^2)=xq(x,\mu^2)-x\overline q(x,\mu^2)$, respectively,  
and the gluon distribution, $xg(x,\mu^2)$ \cite{distribs}.
These distributions are obtained from next-to-leading-order QCD
fits to the structure function data \cite{Seligman} 
using the QCD evolution programs of Duke and Owens \cite{DO}.

To resolve the strange component of the quark sea, the singlet and non-singlet
quark distributions are separated by flavor.
Insofar as isospin is a good symmetry, our experiment is insensitive to the 
exact form of the up and down valence and sea quark distributions,
because the neutrino target is nearly isoscalar. An 
isoscalar correction accounts for the 5.67\% neutron excess in the target. 

The proton valence quark content, $xq_V(x,\mu^2)=xq_{NS}(x,\mu^2)$, is 
parameterized by
\begin{eqnarray}
xq_V(x,\mu^2) & = & xu_V(x,\mu^2) + xd_V(x,\mu^2), \nonumber \\
xd_V(x,\mu^2) & = & A_d(1-x) xu_V(x,\mu^2), 
\end{eqnarray}
where the shape difference for $xd_V(x)$ better fits charged-lepton scattering 
measurements of $F_2^{\rm n}/F_2^{\rm p}$ \cite{nmcf}.
$A_d$ is fixed by demanding that the ratio of the number of $d$ to $u$ valence 
quarks in the proton is 1/2.

The non-strange quark and antiquark components of the sea 
are assumed to be symmetric, so that $x\overline u(x,\mu^2)=x u_S(x,\mu^2)$, 
$x\overline d(x,\mu^2)=x d_S(x,\mu^2)$.
The isoscalar correction is applied assuming 
$x\overline u(x,\mu^2)=x \overline d(x,\mu^2)$.
The strange component of the quark sea is allowed to have a different 
magnitude and shape from the non-strange component.
The strange quark content is set by the parameter 
\begin{equation}
\kappa={\int_0^1[xs(x,\mu^2)+x\overline s(x,\mu^2)]\,dx\over 
\int_0^1[x\overline u(x,\mu^2)+x\overline d(x,\mu^2)]\,dx},
\end{equation}
where $\kappa=1$ would indicate a flavor SU(3) symmetric sea.
The shape of the strange quark distribution 
relates to that of the non-strange sea by a shape parameter $\alpha$, 
where $\alpha=0$ would indicate that the strange sea has the same $x$
dependence as the non-strange component of the quark sea. 
Shape parameters are defined for each of the two fits below. 
In the first fit the strange quark and
antiquark distributions are assumed to be the same; in the second fit
$xs(x,\mu^2)$ and $x\overline s(x,\mu^2)$ are fit separately.

\noindent{\it 4.1 $xs(x,\mu^2)=x\overline s(x,\mu^2)$ fit}

This fit assumes that $xs(x,\mu^2)$ and $x\overline s(x,\mu^2)$ 
are the same. 
The sea quark distributions are 
parameterized by:
\begin{eqnarray}
x\overline q(x,\mu^2) & = & 2 
\left[{x\overline u(x,\mu^2) + 
x\overline d(x,\mu^2)\over 2}\right] + x s(x,\mu^2), \nonumber \\
xs(x,\mu^2) & = & A_s(1-x)^\alpha \left[{x\overline u(x,\mu^2) + 
x\overline d(x,\mu^2)\over 2}\right],
\end{eqnarray}
where $A_s$ is defined in terms of $\kappa$ and $\alpha$.

\begin{figure}[p]
  \centerline{  \hspace{1.3cm}
  \epsfysize=9cm \epsfbox[75 144 550 576]{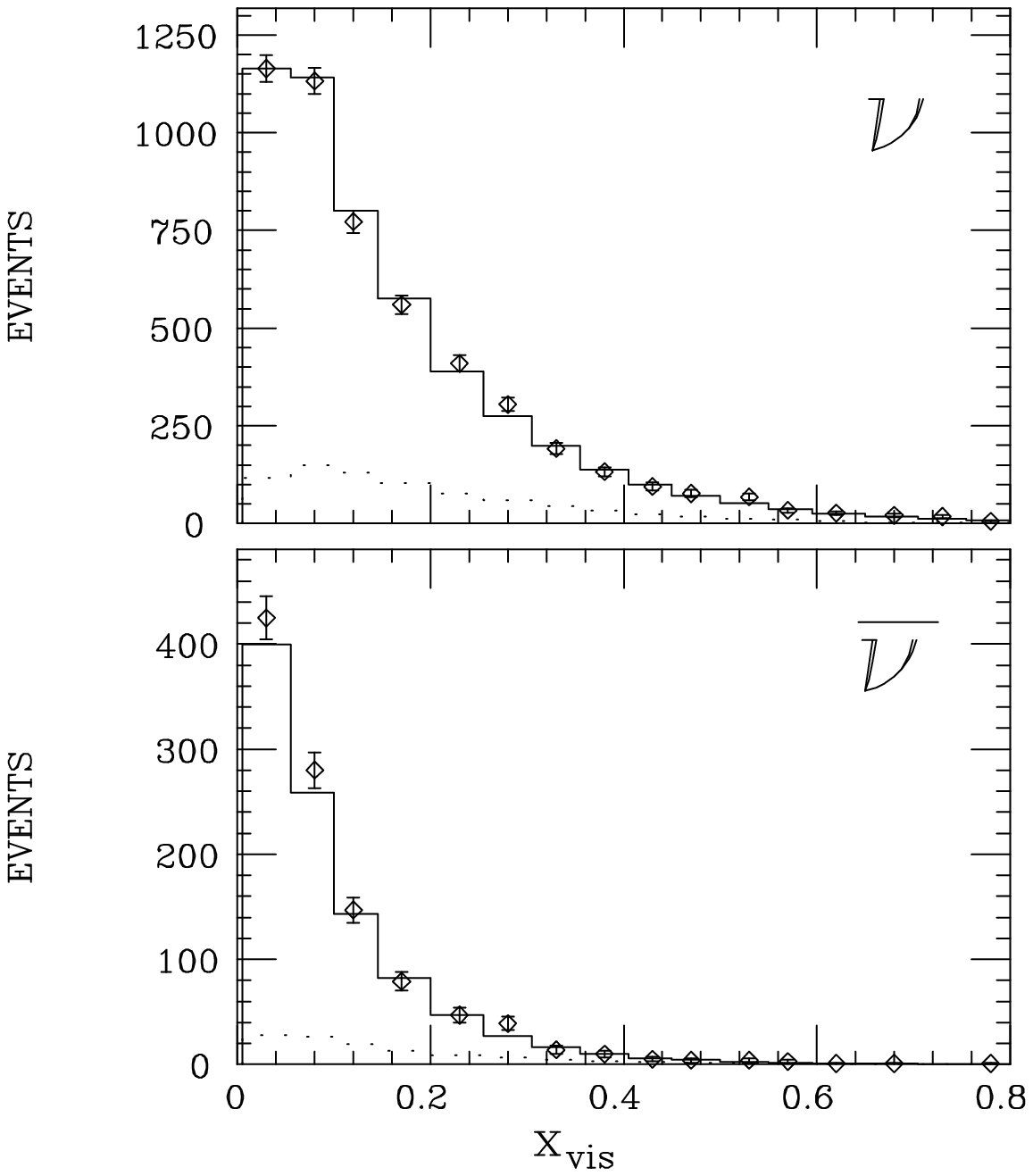} \hspace{-1.5cm}
  \epsfysize=9cm \epsfbox[75 144 550 576]{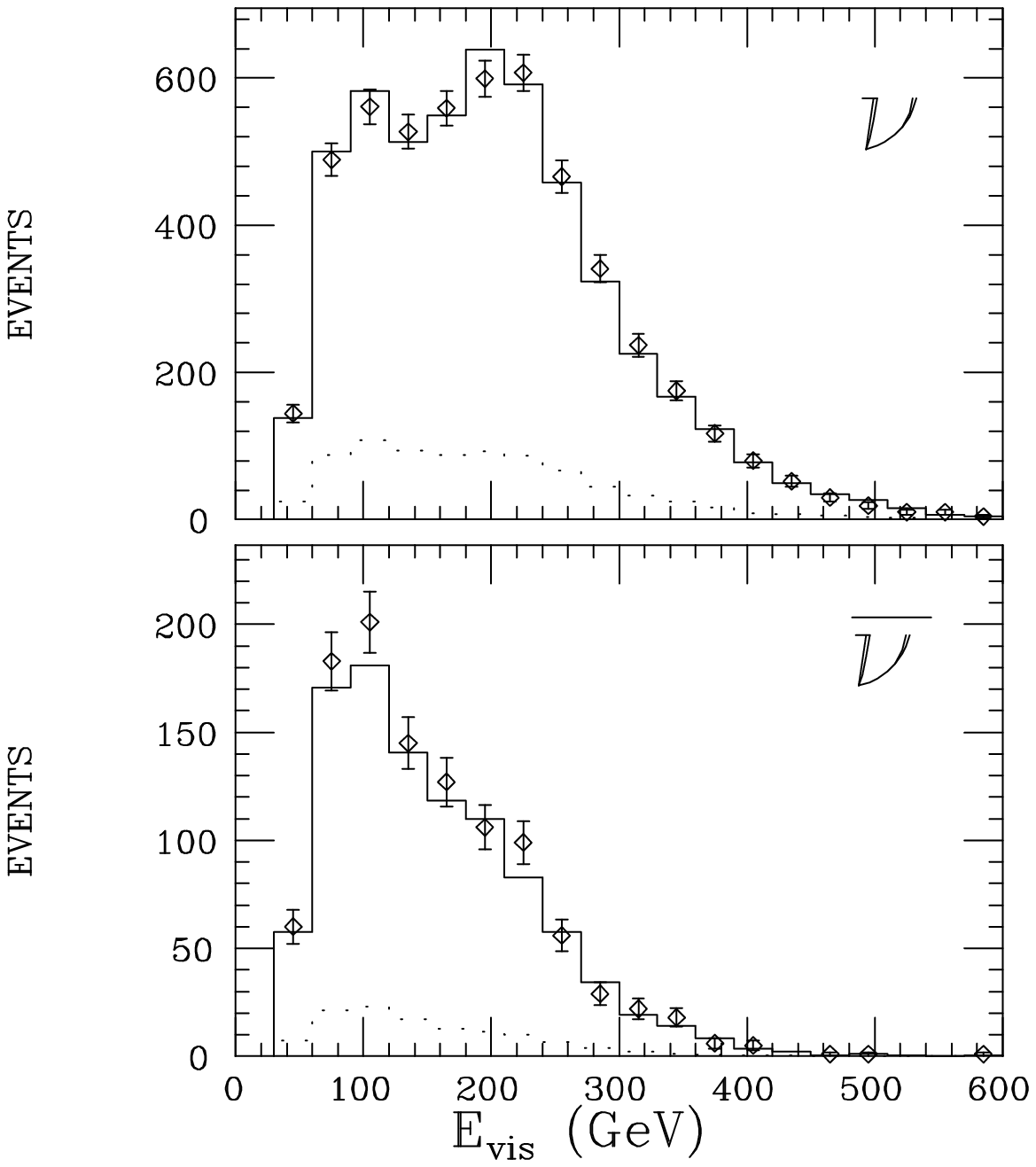}}
  \vspace*{-0.5cm}
  \centerline{
  \epsfysize=9cm \epsfbox[75 144 550 576]{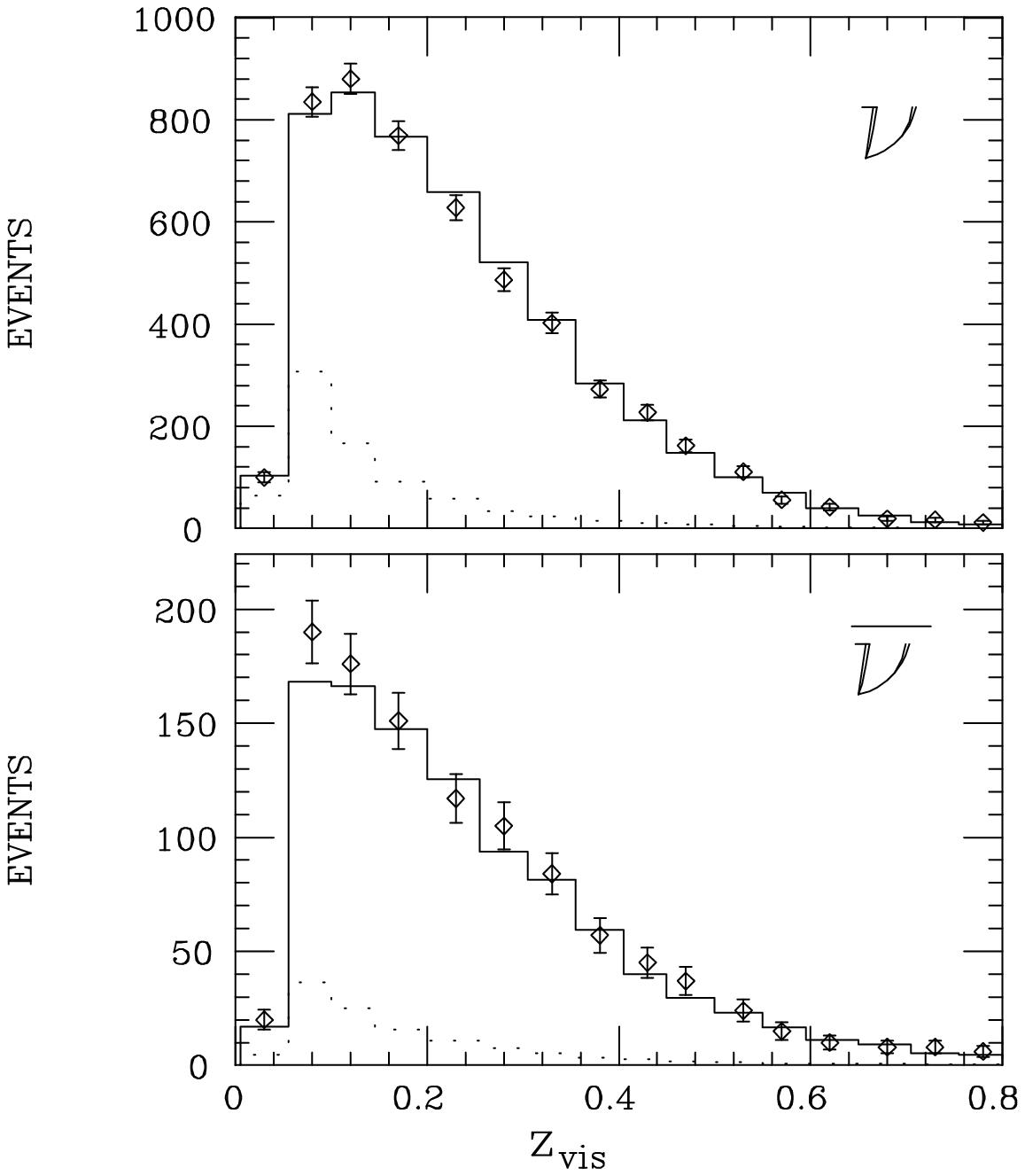}}
   \caption[]{
$x_{\rm vis}$, $E_{\rm vis}$ and $z_{\rm vis}$ distributions 
for $\nu$- and $\overline\nu$-induced dimuon events. Data are given
by the points and the solid histogram is the result of fitting the dimuon
event simulation. The dotted histogram is the background contribution 
to the former from pion and kaon decay.
}
\label{fig:data}
\end{figure}

A $\chi^2$ minimization is performed to find
the strange sea parameters $\kappa$ and $\alpha$, the 
values of $B_c$ and $m_c$, and the fragmentation parameter $\epsilon$, by
fitting to the $x_{\rm vis}$, $E_{\rm vis}$ and $z_{\rm vis}$ distributions of
the dimuon data, projections of which are shown in Figure \ref{fig:data}.
Taking $|V_{cd}|=0.221\pm 0.003$ 
and $|V_{cs}|=0.9743\pm 0.0008$ \cite{pdg} as input values 
and using the Collins-Spiller fragmentation function, 
the extracted NLO parameters 
with their statistical and systematic errors
are presented in the first line of Table \ref{tab:ccfr}. 
In this primary fit, 
the uncertainties due to fragmentation are included in the 
statistical errors of all of the parameters. 
The value of $\chi^2=52.2$ for 65 degrees of freedom suggests excellent 
agreement between the data and the NLO theoretical model. 

Our previous
LO results \cite{sar}, which were found by fitting to the $x_{\rm vis}$ and
$E_{\rm vis}$ distributions of the same data sample and  
using the Peterson fragmentation function \cite{peterson},
 $D(z)= N \{\, z[1-(1/z)-\epsilon_P /(1-z)]^2\,\}^{-1}$ with $\epsilon_P=0.20$,
are listed in the third line
of Table 1. For comparison with these results, Table 1 includes 
the NLO parameters determined using the same fit procedure. The two fits using 
the Peterson
fragmentation function include the uncertainty due to fragmentation in the
systematic errors of all of the parameters.

{\small
\begin{table}[h,b,t]
\begin{center}
\begin{tabular}{|c|c|c|c|c|c|c|}
\hline
\multicolumn{1}{|c|}{}&
  \multicolumn{1}{|c|}{Fragmentation} & 
  \multicolumn{1}{|c|}{$\chi^2/$dof} & 
  \multicolumn{1}{|c|}{$\kappa$} & 
  \multicolumn{1}{|c|}{$\alpha$} & 
  \multicolumn{1}{|c|}{$B_c$} &
  \multicolumn{1}{|c|}{ $m_c$ (${\rm GeV/c^2}$)}  \\
\hline
\multicolumn{1}{|c|}{NLO fit }&
  \multicolumn{1}{|c|}{Collins-Spiller} &
  \multicolumn{1}{|c|}{$ 52.2/65                   $} &
  \multicolumn{1}{|c|}{$ 0.477                     $} &
  \multicolumn{1}{|c|}{$ -0.02                   $} &
  \multicolumn{1}{|c|}{$ 0.1091                   $} &
  \multicolumn{1}{|c|}{$ 1.70                   $}
\\       
\multicolumn{1}{|c|}{}&
  \multicolumn{1}{|c|}{$ \epsilon = 0.81\pm 0.14   $} &
  \multicolumn{1}{|c|}{$                           $} &
  \multicolumn{1}{|c|}{$ ^{+0.046}_{-0.044}\; ^{+0.023}_{-0.024}\;   $} &
  \multicolumn{1}{|c|}{$ ^{+0.60 }_{-0.54 }\; ^{+0.28 }_{-0.26 }\;   $} &
  \multicolumn{1}{|c|}{$ ^{+0.0082}_{-0.0074}\; ^{+0.0063}_{-0.0051}\;$} &
  \multicolumn{1}{|c|}{$ \pm 0.17 \; ^{+0.09 }_{-0.08 }          $}\\
\hline
\multicolumn{1}{|c|}{NLO fit }&
  \multicolumn{1}{|c|}{Peterson       } &
  \multicolumn{1}{|c|}{$ 41.2/46                   $} &
  \multicolumn{1}{|c|}{$ 0.468                     $} &
  \multicolumn{1}{|c|}{$ -0.05                   $} &
  \multicolumn{1}{|c|}{$ 0.1047                   $} &
  \multicolumn{1}{|c|}{$ 1.69                   $}
\\       
\multicolumn{1}{|c|}{}&
  \multicolumn{1}{|c|}{$ \epsilon_P = 0.20\pm 0.04 $} &
  \multicolumn{1}{|c|}{$                           $} &
  \multicolumn{1}{|c|}{$ ^{+0.061}_{-0.046}\; ^{+0.024}_{-0.025}\;   $} &
  \multicolumn{1}{|c|}{$ ^{+0.46 }_{-0.47 }\; ^{+0.28 }_{-0.26 }\;   $} &
  \multicolumn{1}{|c|}{$ \pm 0.0076        \; ^{+0.0065}_{-0.0052}\;$} &
  \multicolumn{1}{|c|}{$ \pm 0.16 \; ^{+0.12 }_{-0.10 }          $}\\
\hline
\multicolumn{1}{|c|}{LO fit }&
  \multicolumn{1}{|c|}{Peterson       } &
  \multicolumn{1}{|c|}{$ 42.5/46                   $} &
  \multicolumn{1}{|c|}{  0.373           } &
  \multicolumn{1}{|c|}{  2.50            } &
  \multicolumn{1}{|c|}{  0.1050          } &      
  \multicolumn{1}{|c|}{  1.31          } \\
\multicolumn{1}{|c|}{ Ref. [6]}&
  \multicolumn{1}{|c|}{$ \epsilon_P = 0.20\pm 0.04 $} &
  \multicolumn{1}{|c|}{$                           $} &
  \multicolumn{1}{|c|}{$ ^{+0.048}_{-0.041} \pm 0.018 $} &
  \multicolumn{1}{|c|}{$ ^{+0.60}_{-0.55}\: ^{+0.36}_{-0.25}  $} &
  \multicolumn{1}{|c|}{$      \pm 0.007 \pm 0.005 $} &
  \multicolumn{1}{|c|}{$ ^{+0.20}_{-0.22}\: ^{+0.12}_{-0.11}  $}\\
\hline
\end{tabular}
\end{center} 
\vspace{-2ex}
 \caption[]{
Next-to-leading-order and leading-order fit results, assuming
$xs(x)=x\overline s(x)$. Errors are statistical and systematic, except that the
errors on the fragmentation parameters are statistical only.}
 \label{tab:ccfr}
\end{table} }   

Estimates of the systematic uncertainties are obtained by varying 
model parameters
within errors and are itemized in Table \ref{tab:err}. 

{\renewcommand{\arraystretch}{1.2}
\begin{table}[h,b,t]
\begin{center}
\begin{tabular}{|c|c|c|c|c|}
\hline
  \multicolumn{1}{|c|}{source of uncertainty} & 
  \multicolumn{1}{|c|}{$\kappa$} & 
  \multicolumn{1}{|c|}{$\alpha$} & 
  \multicolumn{1}{|c|}{$B_c$} &
  \multicolumn{1}{|c|}{$m_c$}  \\
\hline
  \multicolumn{1}{|c|}{$\pi/K$ background          } &
  \multicolumn{1}{|c|}{$ \;^{+\; 0.0010 }_{-\; 0.0022  }$   } &
  \multicolumn{1}{|c|}{$ \;^{+\; 0.150}_{-\; 0.160}$        } &
  \multicolumn{1}{|c|}{$ \;^{-\; 0.0031 }_{+\; 0.0027  }$    } &
  \multicolumn{1}{|c|}{$ \;^{+\; 0.006  }_{-\; 0.003  }$       } \\
\hline
  \multicolumn{1}{|c|}{energy scale          } &
  \multicolumn{1}{|c|}{$ \;^{-\; 0.0051 }_{+\; 0.0078  }$   } &
  \multicolumn{1}{|c|}{$ \;^{-\; 0.048}_{+\; 0.059}$        } &
  \multicolumn{1}{|c|}{$ \;^{+\; 0.0028 }_{-\; 0.0002  }$    } &
  \multicolumn{1}{|c|}{$ \;^{+\; 0.031  }_{-\; 0.057  }$       } \\
\hline
  \multicolumn{1}{|c|}{relative calibration       } &
  \multicolumn{1}{|c|}{$ \;^{-\; 0.0059 }_{+\; 0.0020  }$   } &
  \multicolumn{1}{|c|}{$ \;^{+\; 0.115}_{-\; 0.077}$        } &
  \multicolumn{1}{|c|}{$ \;^{+\; 0.0013 }_{-\; 0.0002  }$    } &
  \multicolumn{1}{|c|}{$ \;^{+\; 0.031  }_{-\; 0.007  }$       } \\
\hline
  \multicolumn{1}{|c|}{ detection efficiency $\pm 1\%$      } &
  \multicolumn{1}{|c|}{$ \;^{-\; 0.0070 }_{+\; 0.0021  }$   } &
  \multicolumn{1}{|c|}{$ \;^{+\; 0.044}_{-\; 0.003}$        } &
  \multicolumn{1}{|c|}{$ \;^{+\; 0.0022 }_{-\; 0.0012  }$    } &
  \multicolumn{1}{|c|}{$ \;^{-\; 0.007  }_{+\; 0.013  }$       } \\
\hline
  \multicolumn{1}{|c|}{$\nu-\overline\nu$ mis-id   } &
  \multicolumn{1}{|c|}{$ \;^{+\; 0.0014 }_{-\; 0.0032  }$   } &
  \multicolumn{1}{|c|}{$ \;^{-\; 0.042}_{+\;0.006 }$        } &
  \multicolumn{1}{|c|}{$ \;^{+\; 0.0009 }_{-\;0.0014   }$    } &
  \multicolumn{1}{|c|}{$ \;^{+\; 0.036  }_{-\; 0.030  }$       } \\
\hline
  \multicolumn{1}{|c|}{$F_2$ and $xF_3$            } &
  \multicolumn{1}{|c|}{$ \;^{+\; 0.0143 }_{-\;0.0068   }$   } &
  \multicolumn{1}{|c|}{$ \;^{+\; 0.084}_{-\;0.114 }$        } &
  \multicolumn{1}{|c|}{$ \;^{+\; 0.0024 }_{-\;0.0019   }$    } &
  \multicolumn{1}{|c|}{$ \;^{+\; 0.053  }_{-\;0.017   }$       } \\
\hline
  \multicolumn{1}{|c|}{$xq_{SI}(x)$, $xq_{NS}(x)$ and $xg(x)$ } &
  \multicolumn{1}{|c|}{$ \;^{+\; 0.0110 }_{-\; 0.0110}$ } &
  \multicolumn{1}{|c|}{$ \;^{+\;0.180  }_{-\;0.180 }    $   } &
  \multicolumn{1}{|c|}{$ \;^{+\;0.0009}_{-\;0.0009 }    $} &
  \multicolumn{1}{|c|}{$ \;^{+\;0.038   }_{-\;0.038  }  $} \\
\hline
  \multicolumn{1}{|c|}{$|V_{cs}|$ and $|V_{cd}|$   } &
  \multicolumn{1}{|c|}{$ \;^{+\; 0.0095 }_{-\; 0.0099  }$   } &
  \multicolumn{1}{|c|}{$ \;^{-\; 0.024}_{+\; 0.031}$        } &
  \multicolumn{1}{|c|}{$ \;^{-\; 0.0028 }_{+\;0.0032   }$    } &
  \multicolumn{1}{|c|}{$ \;^{-\; 0.011  }_{+\;0.014   }$       } \\
\hline
  \multicolumn{1}{|c|}{ t--u channel accep. corr.             } &
  \multicolumn{1}{|c|}{$ \;^{-\; 0.0135 }_{+\; 0.0052  }$   } &
  \multicolumn{1}{|c|}{$ \;^{-\; 0.009}_{+\; 0.062}$        } &
  \multicolumn{1}{|c|}{$ \;^{-\; 0.0002 }_{+\; 0.0010  }$    } &
  \multicolumn{1}{|c|}{$ \;^{-\; 0.020  }_{+\; 0.008  }$       } \\
\hline
\end{tabular}
\end{center}
\vspace{-2ex}
 \caption[]{Sources of systematic uncertainty in the determination of the fit
parameters.                                      }
 \label{tab:err}
\end{table}  }  

Parton distributions are defined to a given order and scheme in QCD.
Therefore, the magnitude of a given parton distribution 
differs between leading-order and next-to-leading-order.
At NLO, the nucleon strange quark content is found to be 
$\kappa=0.477\:^{+0.051}_{-0.050}$, indicating that the sea
is not SU(3) symmetric---qualitatively the same result as from 
the LO analysis. The strange quark content may alternatively be given by
\begin{equation}
\eta={\int_0^1[xs(x,\mu^2)+x\overline s(x,\mu^2)]\,dx\over 
\int_0^1[x u(x,\mu^2)+x d(x,\mu^2)]\,dx},
\end{equation}
so that by comparing to the total non-strange quark content, $\eta$ is less
sensitive to changes in the determination of the sea quark content alone.
At $\mu^2=22.2$ $\rm GeV^2/c^2$ the ratio of antiquarks to quarks in the
nucleon at NLO is found to be $\int dx x\overline q(x,\mu^2)/\int 
dx xq(x,\mu^2) =\overline Q/Q= 0.245 \pm 0.005$ and thereby 
the strange quark content with respect to the non-strange quarks is 
\begin{equation}
\eta = 0.099 \pm 0.008 \pm 0.004 \;^{-\; 0.003}_{+\; 0.006}.
\end{equation}

\begin{figure}[b,t,h,p]
\begin{minipage}[t]{78mm}
   \epsfysize=5cm \epsfbox[36 230 540 576]{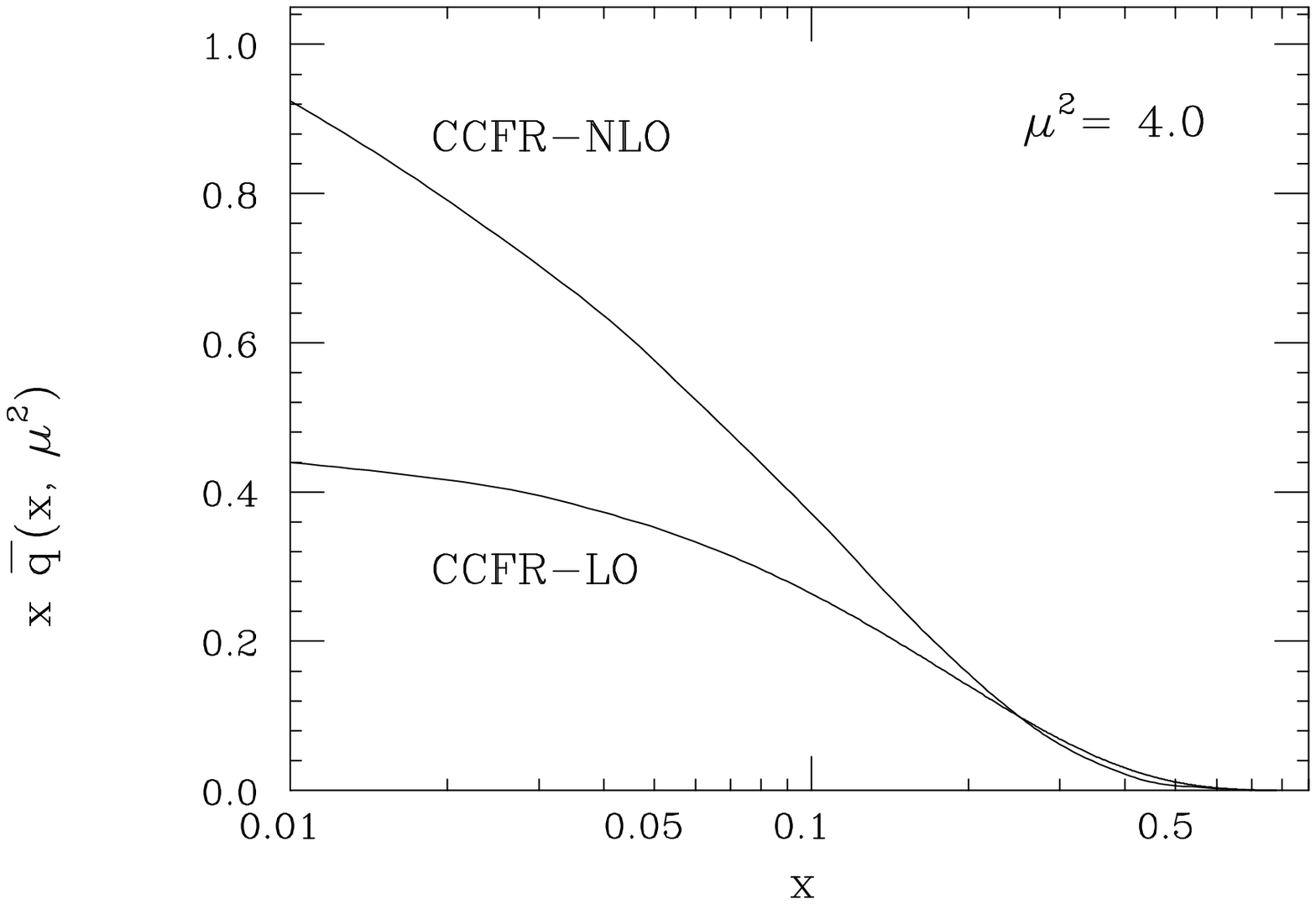}
   \caption[]{
The quark sea distribution $x\overline q(x,\mu^2=4.0\: {\rm GeV^2/c^2})$ 
determined at next-to-leading order and leading order.
   }
\label{fig:qbar}
\end{minipage}
\hspace{\fill}
\begin{minipage}[t]{77mm}
    \epsfysize=5cm \epsfbox[36 230 540 576]{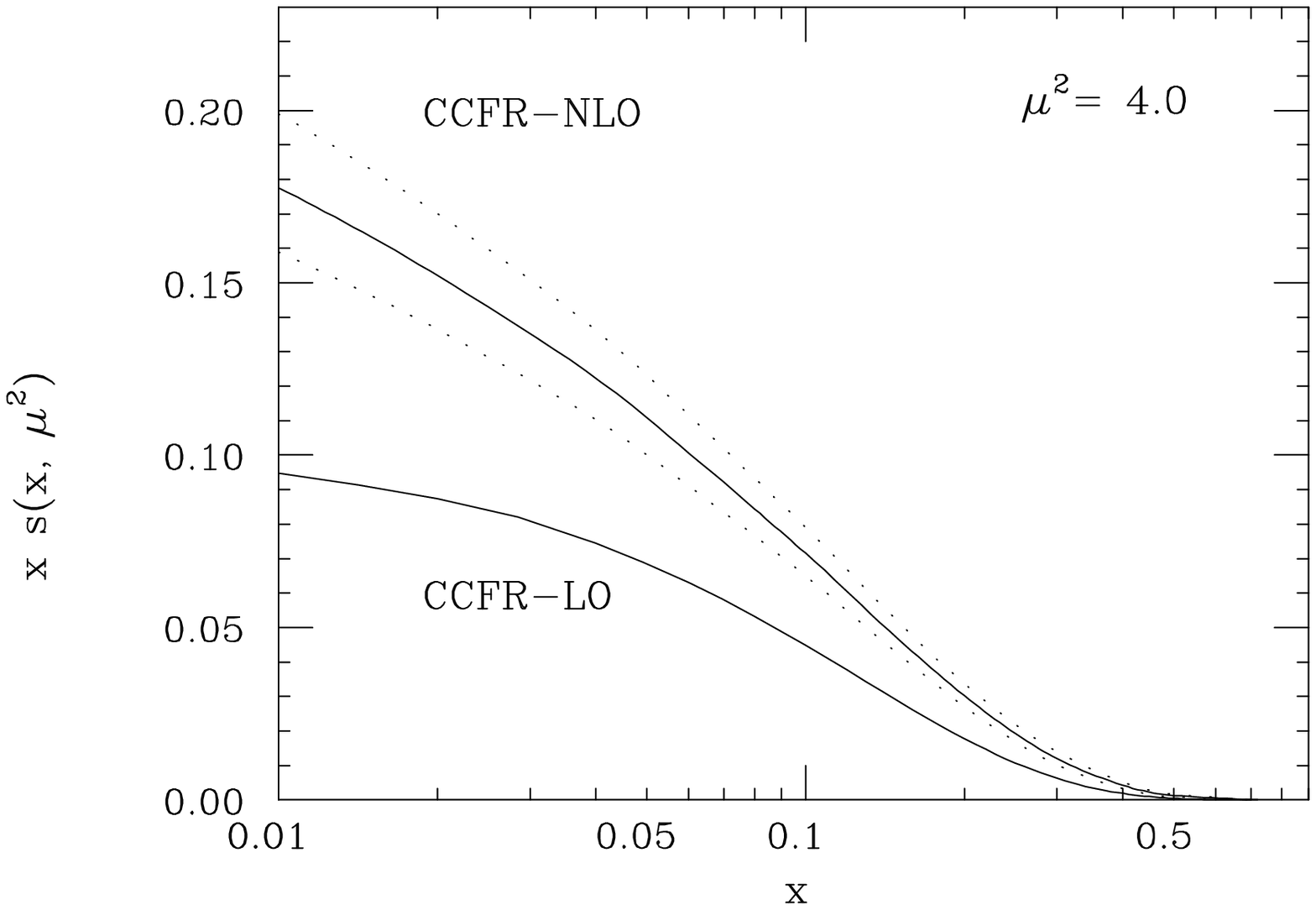}
    \caption[]{
The strange quark distribution $x s(x,\mu^2=4.0\: {\rm GeV^2/c^2})$ 
determined at next-to-leading order (described in section 4.1) 
and leading order. The band around the 
NLO curve indicates the $\pm 1\sigma$ uncertainty in the distribution.
     }
\label{fig:ss}
\end{minipage}
\end{figure}

Since a nonzero value of $\alpha$ would indicate 
a shape difference between $x\overline
q(x)$ and $xs(x)$, the value $\alpha=-0.02\;^{+\;0.66}_{-\;0.60} $ 
indicates no shape difference at NLO. At leading order, we 
find the strange quarks softer than the overall quark sea by a
factor $(1-x)^\alpha$ with $\alpha= 2.5\pm 0.7$.
The difference in $\alpha$ between NLO and LO is attributable to the
NLO $x\overline q(x)$ 
being softer than its LO counterpart, as shown in Figure \ref{fig:qbar}.
Figure \ref{fig:ss} shows the NLO and LO $xs(x)$, again indicating that the
NLO distribution is larger in magnitude and softer than its LO counterpart.

The strange quark distribution from the $xs(x,\mu^2)=x\overline s(x,\mu^2)$ fit
is tabulated in Table \ref{tab:strange} for a few values of $x$ and $\mu^2$ and
is plotted at $\mu^2=4$ $\rm GeV^2/c^2$ in Figure
\ref{fig:ss}. The distribution can be parameterized by a function of the form
$a\,(1-x)^b \, x^{-c}$. The values of the coefficients $a$, $b$ and $c$ are
tabulated in Table \ref{tab:param}.

{\renewcommand{\arraystretch}{1.4}
\begin{table}[h,b,t]
\begin{center}
\begin{tabular}{|c|c|c|}
\hline
$\mu^2$ ($\rm GeV^2/c^2$) &  $x$  & $x s(x,\mu^2)$    \\ \hline
%output from pdfout2.f, the strange sea,  sfac and salpha = 0.4770 -0.0200
 1.0 & 0.01 & $ 0.126  \pm 0.012 \pm 0.006 \;^{+\;0.008}_{-\;0.004}$\\
     & 0.05 & $ 0.097  \pm 0.008 \pm 0.004 \;^{+\;0.006}_{-\;0.003}$\\
     & 0.10 & $ 0.068  \pm 0.005 \pm 0.003 \;^{+\;0.004}_{-\;0.002}$\\
     & 0.20 & $ 0.032  \pm 0.003 \pm 0.001 \;^{+\;0.002}_{-\;0.001}$\\ \hline
 4.0 & 0.01 & $ 0.178  \pm 0.016 \pm 0.008 \;^{+\;0.011}_{-\;0.005}$\\
     & 0.05 & $ 0.111  \pm 0.009 \pm 0.005 \;^{+\;0.007}_{-\;0.003}$\\
     & 0.10 & $ 0.072  \pm 0.005 \pm 0.003 \;^{+\;0.004}_{-\;0.002}$\\
     & 0.20 & $ 0.030  \pm 0.003 \pm 0.001 \;^{+\;0.002}_{-\;0.001}$\\ \hline
20.0 & 0.01 & $ 0.229  \pm 0.020 \pm 0.010 \;^{+\;0.014}_{-\;0.007}$\\
     & 0.05 & $ 0.122  \pm 0.010 \pm 0.005 \;^{+\;0.007}_{-\;0.004}$\\
     & 0.10 & $ 0.073  \pm 0.006 \pm 0.003 \;^{+\;0.004}_{-\;0.002}$\\
     & 0.20 & $ 0.028  \pm 0.003 \pm 0.001 \;^{+\;0.002}_{-\;0.001}$\\ \hline
100.0& 0.01 & $ 0.271  \pm 0.024 \pm 0.012 \;^{+\;0.017}_{-\;0.008}$\\
     & 0.05 & $ 0.128  \pm 0.010 \pm 0.005 \;^{+\;0.008}_{-\;0.004}$\\
     & 0.10 & $ 0.072  \pm 0.006 \pm 0.003 \;^{+\;0.004}_{-\;0.002}$\\
     & 0.20 & $ 0.026  \pm 0.003 \pm 0.001 \;^{+\;0.002}_{-\;0.001}$\\ \hline
\end{tabular}
\end{center}
\vspace{-2ex}
 \caption[]{Values of $xs(x,\mu^2)$, defined at NLO using the $\overline{\rm MS}$
renormalization scheme, from the $xs(x,\mu^2)=x\overline s(x,\mu^2)$ fit. The
first error is statistical, the second is experimental systematic and the third
is due to QCD scale uncertainty.}
 \label{tab:strange}
\end{table} }   

{\renewcommand{\arraystretch}{1.4}
\begin{table}[h,b,t]
\begin{center}
\begin{tabular}{|c|ccc|}
\hline
$\mu^2$ $\rm GeV^2/c^2$ &  $a$  & $b$ & $c$   \\ \hline
 1.0 & 0.135 &   6.48  &   0.000 \\
10.0 & 0.117 &   6.67  &   0.096 \\
20.0 & 0.107 &   6.98  &   0.164 \\
100.0& 0.100 &   7.28  &   0.210 \\ \hline
\end{tabular}
\end{center}
\vspace{-2ex}
 \caption[]{The coefficients $a$, $b$ and $c$ from the parameterization 
of $xs(x,\mu^2)$ 
using the form $a(1-x)^bx^{-c}$, as described in the text. 
} \label{tab:param}
\end{table} }   

The charm quark mass parameter from the NLO fit is 
$1.70 \pm 0.19$ GeV/c$^2$, which differs from the leading-order
result, indicating the marked dependence of $m_c$ on the order 
to which the analysis is done. The NLO value of $m_c$ can be more consistently
compared 
with measurements derived from other processes involving similar higher-order
perturbative QCD calculations.
A photon-gluon-fusion analysis of photoproduction data
finds $m_c=1.74\:^{+0.13}_{-0.18}$ \cite{photo}. 

The values of $B_c$ from the NLO and LO fits are consistent, providing a good 
check of the fit procedure and the physics model.
This value is constrained by neutrino charm production at $x>0.3$ where the 
valence $d$ quark contribution dominates, and thus is independent of the
strange and other sea quark distributions.
The value of $B_c=0.109 \:^{+0.010}_{-0.007}$ agrees 
with an indirect determination, 
$B_c^{\rm I}=0.099\pm0.012$, which is described in section 5. 

\begin{figure}[h,b,t]
\begin{minipage}[t]{78mm}
   \epsfysize=9cm \epsfbox[72 144 518 627]{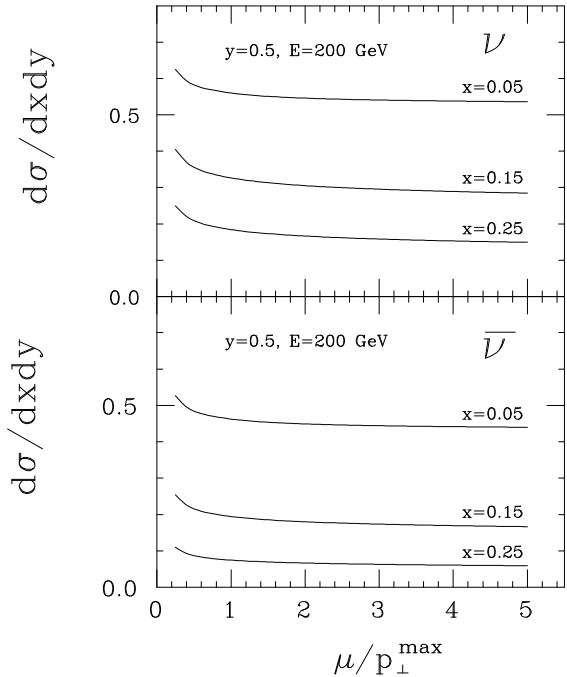}
   \caption[]{
   The $\mu^2$ scale dependence of the differential cross
section for neutrino and antineutrino production of charm, where
$\mu^2$ identifies the factorization and renormalization scales.  
The scale $\mu$ on the abscissa
is in units of $p_\perp^{\rm max}$. For $E=200$ GeV and $y=0.5$, the $x=0.05$, 
0.15, 0.25 lines correspond to $p_\perp^{\rm max}=6.6$, 6.2, 5.8 GeV/c,
respectively. 
   }
\label{fig:scale}
\end{minipage}
\end{figure}

As with all applications of perturbative QCD, a theoretical uncertainty is
associated with the choice of factorization and renormalization scales. Some
scale dependence is unavoidable for any calculation done to finite 
order in $\alpha_S$. 
The $\mu^2$ scale is interpreted as setting the 
boundary between the collinear and noncollinear regions of the $p_\perp$
integration over the final states.  
Therefore, a scale proportional to $p_\perp^{\rm max}$
is suggested by the authors of Ref. \cite{AOT,scale}. 
Figure \ref{fig:scale} shows the scale dependence of
the differential cross section, where the abscissa is in units 
of $p_\perp^{\rm max}$. 
The scale dependence is
weak for $\mu$ values above one unit of $p_\perp^{\rm max}$, and there 
is a stronger scale dependence when $\mu$ is below this value.
We choose $\mu=2p_\perp^{\rm max}$ 
and find the scale uncertainty by varying $\mu$
between $p_\perp^{\rm max}$ and $3p_\perp^{\rm max}$. 
Fit results with
various choices of the common factorization and renormalization scale 
are presented in
Table \ref{tab:scale}. The $\chi^2$ values for these fits---all with 65
degrees of freedom---indicate that the data favor $\mu^2$ scales with smaller
magnitudes. 
It should be noted that the values of the fit parameters are fairly insensitive
to the choice of scale.

{\renewcommand{\arraystretch}{1.5}
\begin{table}[b,t,h,p]
\begin{center}
\begin{tabular}{|c|c|c|c|c|c|}
\hline
  \multicolumn{1}{|c|}{choice of scale, $\mu^2$} & 
  \multicolumn{1}{|c|}{$\chi^2$} & 
  \multicolumn{1}{|c|}{$\kappa$} & 
  \multicolumn{1}{|c|}{$\alpha$} & 
  \multicolumn{1}{|c|}{$B_c$} &
  \multicolumn{1}{|c|}{ $m_c$ $({\rm GeV/c^2})$ } \\
\hline\hline
  \multicolumn{1}{|c|}{$(\: p_\perp^{\rm max})^2$      } &
  \multicolumn{1}{|c|}{  50.4                       } &
  \multicolumn{1}{|c|}{  0.513                      } &
  \multicolumn{1}{|c|}{  0.18                       } &
  \multicolumn{1}{|c|}{  0.0987                     } &
  \multicolumn{1}{|c|}{  1.71                      } \\
\hline
  \multicolumn{1}{|c|}{$(2\: p_\perp^{\rm max})^2$      } &
  \multicolumn{1}{|c|}{  52.2                       } &
  \multicolumn{1}{|c|}{  0.477                      } &
  \multicolumn{1}{|c|}{$-0.02$                      } &
  \multicolumn{1}{|c|}{  0.1091                     } &
  \multicolumn{1}{|c|}{  1.70                      } \\
\hline
  \multicolumn{1}{|c|}{$(3\: p_\perp^{\rm max})^2$     } &
  \multicolumn{1}{|c|}{  54.4                       } &
  \multicolumn{1}{|c|}{  0.460                      } &
  \multicolumn{1}{|c|}{$-0.10$                      } &
  \multicolumn{1}{|c|}{  0.1142                     } &
  \multicolumn{1}{|c|}{  1.68                      } \\
\hline\hline
  \multicolumn{1}{|c|}{ $Q^2$        } &
  \multicolumn{1}{|c|}{  51.7                       } &
  \multicolumn{1}{|c|}{  0.423                      } &
  \multicolumn{1}{|c|}{ $-0.37$                     } &
  \multicolumn{1}{|c|}{  0.1074                     } &
  \multicolumn{1}{|c|}{  1.80                      } \\
\hline
  \multicolumn{1}{|c|}{$(2Q)^2$  } &
  \multicolumn{1}{|c|}{  56.1                       } &
  \multicolumn{1}{|c|}{  0.410                      } &
  \multicolumn{1}{|c|}{ $-0.46$                      } &
  \multicolumn{1}{|c|}{  0.1159                     } &
  \multicolumn{1}{|c|}{  1.71                      } \\
\hline
  \multicolumn{1}{|c|}{$(3Q)^2$            } &
  \multicolumn{1}{|c|}{  59.4                       } &
  \multicolumn{1}{|c|}{  0.408                      } &
  \multicolumn{1}{|c|}{$-0.54$                      } &
  \multicolumn{1}{|c|}{  0.1206                     } &
  \multicolumn{1}{|c|}{  1.73                      } \\
\hline\hline
  \multicolumn{1}{|c|}{$Q^2+m_c^2$ } &
  \multicolumn{1}{|c|}{  52.5                       } &
  \multicolumn{1}{|c|}{  0.421                      } &
  \multicolumn{1}{|c|}{$-0.03$                      } &
  \multicolumn{1}{|c|}{  0.1066                     } &
  \multicolumn{1}{|c|}{  1.65                      } \\
\hline
  \multicolumn{1}{|c|}{$4\:(Q^2+m_c^2)$} &
  \multicolumn{1}{|c|}{  57.1                       } &
  \multicolumn{1}{|c|}{  0.409                      } &
  \multicolumn{1}{|c|}{$-0.16$                      } &
  \multicolumn{1}{|c|}{  0.1154                     } &
  \multicolumn{1}{|c|}{  1.64                      } \\
\hline\hline
  \multicolumn{1}{|c|}{$Q^2+(2m_c)^2$         } &
  \multicolumn{1}{|c|}{  52.8                       } &
  \multicolumn{1}{|c|}{  0.428                      } &
  \multicolumn{1}{|c|}{  0.00                       } &
  \multicolumn{1}{|c|}{  0.1068                     } &
  \multicolumn{1}{|c|}{  1.62                      } \\
\hline
  \multicolumn{1}{|c|}{$4\: [Q^2+(2m_c)^2]$   } &
  \multicolumn{1}{|c|}{  57.3                       } &
  \multicolumn{1}{|c|}{  0.415                      } &
  \multicolumn{1}{|c|}{$-0.15$                      } &
  \multicolumn{1}{|c|}{  0.1161                     } &
  \multicolumn{1}{|c|}{  1.63                      } \\
\hline
\end{tabular}
\end{center}
\vspace{-2ex}
 \caption[]{Central values of the fit parameters for various choices of the QCD 
scale $\mu^2$. Each fit contains 65 degrees of freedom.}
 \label{tab:scale}
\end{table}  }  

\pagebreak
\noindent{\it 4.2 $xs(x)\neq x\overline s(x)$ fit}

Theoretical work has explored the possibility 
that the nucleon contains a sizable heavy quark component at
moderate $x$---the possibility of so-called intrinsic heavy quark states within
the nucleon \cite{brodsky}. 
Postulating intrinsic strange quark states leads to the prediction that the 
$s$ quark momentum distribution will be harder than the
$\overline s$ quark distribution \cite{burkardt}. 
We explore this possibility by performing a fit in which the momentum 
distributions of the
$s$ and $\overline s$ quarks are allowed to be different. 
For this study the sea quark distributions are 
parameterized by:
\begin{eqnarray}
x\overline q(x,\mu^2) & = & 2\left( {x\overline u(x,\mu^2) + 
x\overline d(x,\mu^2)\over 2} \right)
+ {xs(x,\mu^2)+x\overline s(x,\mu^2)\over 2},\nonumber \\
 xs(x,\mu^2) & = & A_s(1-x)^\alpha 
\left[{x\overline u(x,\mu^2) 
+x\overline d(x,\mu^2)\over 2}\right],\nonumber \\
 x\overline s(x,\mu^2) & = & A_s^\prime(1-x)^{\alpha^\prime}
\left[{x\overline u(x,\mu^2) 
+x\overline d(x,\mu^2)\over 2}\right].
\end{eqnarray}
The $s$ and $\overline s$ are constrained to have the same number
\begin{equation}
\int_0^1s(x,\mu^2)\,dx =  \int_0^1\overline s(x,\mu^2)\,dx.
\end{equation}
$A_s$ and $A_s^\prime$ are defined in terms of $\kappa$, $\alpha$ and
$\alpha^\prime$.

In order to reduce the number of free parameters, this fit constrains the
average charmed hadron branching ratio to the value obtained from other
measurements, 
$B_c^{\rm I}=0.099\pm0.012$ (see Section 5.). 
We fit for four parameters: the strange quark parameters 
$\kappa$, $\alpha$, and $\Delta\alpha=\alpha-\alpha^\prime$ and 
the charm quark mass $m_c$. The result is:
\begin{eqnarray}
 \kappa & = & 0.536 \pm 0.030 \pm 0.036 
 \;^{-\; 0.064}_{+\; 0.098} \pm 0.009,\nonumber \\
 \alpha & = & -0.78 \pm 0.40 \pm 0.56 \pm 0.98 \pm 0.50,\nonumber \\
 \Delta\alpha & = & -0.46 \pm 0.42 \pm 0.36 \pm 0.65 \pm 0.17,\nonumber \\
 m_c & = & 1.66 \pm 0.16 \pm 0.07 \;^{+\; 0.04}_{-\; 0.01}
\pm 0.01 \; {\rm GeV/c^2}, 
\end{eqnarray}
where the first error is statistical, the second is systematic, 
the third is due to the uncertainty in $B_c^{\rm I}$,
and the fourth
is the error due to $\mu^2$ scale uncertainty.

The value of $\Delta\alpha= -0.46\pm 0.85 \pm 0.17$ indicates that 
the momentum distributions of $s$ and $\overline s$ are consistent and the 
difference in
the two distributions is limited to
$-1.9 < \Delta\alpha < 1.0$ 
at the 90\% confidence level.
This is the first quantitative comparison 
of the components of the $s$ and $\overline s$ quark sea.

We also checked the assumption that the same average semileptonic branching 
ratio applies to the $\nu$- and $\overline\nu$-induced samples. A two 
parameter fit finds the branching ratio of $\nu$-induced events $B_c=0.1147\pm
0.0056$, and $\Delta B_c=B_c-B_c^\prime=0.011\pm 0.011$, where $B_c^\prime$
is the branching ratio for $\overline\nu$-induced events and the errors are 
statistical only. The result indicates that there is no significant difference
in the semileptonic decays of charmed particles and antiparticles at 
these energies. 

\noindent{\bf 5. $|V_{cd}|$ measurement}

If the CKM matrix elements are not assumed, then the four
parameter NLO fit in section 4.1 is performed by fitting $\alpha$, $m_c$ and
the following products:
\begin{eqnarray}
|V_{cd}|^2B_c & = & (5.34\;^{+\; 0.38}_{-\; 0.39}\;^{+\; 0.27}_{-\; 0.21} 
\;^{+\; 0.25}_{-\; 0.51})\times 10^{-3},\nonumber \\
{\kappa\over\kappa +2} |V_{cs}|^2B_c & = & 
(2.00 \pm 0.10 \;^{+\; 0.07}_{-\; 0.05}
\;^{+\; 0.06}_{-\; 0.14})\times 10^{-2}.
\end{eqnarray}

These combinations can be used to extract $|V_{cd}|^2$ and $\kappa |V_{cs}|^2$
when $B_c$ is determined from other data. $B_c$ is
determined by combining the charmed particle semileptonic branching ratios
measured at $e^+e^-$ colliders \cite{pdg} with
the neutrino-production fractions measured by 
the Fermilab E531 neutrino-emulsion experiment \cite{e531}. 
Using an $E_{\rm vis}>30$ GeV cut,
E531 determined the following production fractions:   
$52\pm 6$\% $D^0$, $42\pm 6$\% $D^+$, 
$1\pm 2$\% $D_s^+$, and $5\pm 3$\% $\Lambda_c^+$.
In the E531 analysis, events that could not be unambiguously identified as
$D^+$ or $D_s^+$ were all categorized as $D^+$ events. 
To remove this small bias, 
a re-analysis was performed that
included updated values of the charmed hadron lifetimes \cite{bolton}.    
This re-analysis finds the following production fractions
with an $E_{\rm vis}>30$ GeV cut:
$60\pm 6$\% $D^0$, $26\pm 6$\% $D^+$, 
$7\pm 5$\% $D_s^+$, and $7\pm 4$\% $\Lambda_c^+$.
These production fractions are consistent with those measured by $e^+e^-$
experiments \cite{cleo}.

We find $B_c^{\rm I}=0.099\pm0.012$ and extract the value of the 
CKM matrix element
\begin{equation}
|V_{cd}|=0.232\;^{+\; 0.018}_{-\; 0.020},
\end{equation}
where the error indicates all sources of uncertainty, including the $\mu^2$
scale uncertainty.
It compares very well with
the PDG value, $|V_{cd}|=0.221\pm 0.003$, which is determined 
from measurements of the other
matrix elements and the unitarity constraint on the CKM matrix assuming three
generations.
A measurement of $|V_{cs}|$ will be possible when an independent 
measurement of the strange sea content is available.

\noindent{\bf 6. Summary}

We have perfomed the first NLO QCD analysis of neutrino charm production and
have measured 
the nucleon strange quark distribution and the electroweak
parameters $m_c$ and $|V_{cd}|$.
We find:
\begin{eqnarray}
m_c & = & 1.70 \pm 0.19 \pm 0.02 \nonumber \\
\kappa&=&0.477 \;^{+\; 0.051}_{-\; 0.050} \;^{-\; 0.017}_{+\; 0.036}\nonumber \\
\alpha& =& -0.02 \;^{+\; 0.66}_{-\; 0.60} \;^{+\; 0.08}_{-\; 0.20}
\end{eqnarray}
where $\kappa$ 
is the strange quark content with respect to the non-strange sea 
and $\alpha$ 
indicates a shape difference between the strange and non-strange sea,
$xs(x)\propto (1-x)^\alpha\; [x\overline u(x) + x\overline d(x)]/2$. 
The first error combines statistical and systematic errors in quadrature and
the second is the uncertainty due to QCD $\mu^2$ scale.
This value of $\kappa$ indicates that the quark sea 
is not flavor SU(3) symmetric. The relative shape parameter $\alpha$ is 
consistent with zero, indicating that 
there is no shape
difference between the strange and non-strange components of the sea.
The value of $m_c$ obtained from the NLO analysis is consistent with that found
in other processes.

Using an externally determined production weighted
charmed hadron branching ratio, $B_c^{\rm I}=0.099\pm0.012$, we measure 
the CKM matrix element
\begin{equation}
|V_{cd}|=0.232\;^{+\; 0.018}_{-\; 0.020}.
\end{equation}

We have also studied the possibility of 
a shape difference between the $xs(x)$ and $x\overline
s(x)$ distributions. We find that
\begin{equation}
 \Delta\alpha=\alpha-\alpha^\prime=-0.46\pm 0.85 \pm 0.17, 
\end{equation}
where $\alpha$ and $\alpha^\prime$ are shape parameters for $xs(x)$ and 
$x\overline s(x)$, 
indicating no shape difference between the components
of the strange quark sea. A shape difference 
is limited to $-1.9 < \alpha-\alpha^\prime <1.0$ at 90\% confidence level.

The strange sea can also be inferred from
a comparison of charged lepton and neutrino structure functions. To
leading-order the lepton and neutrino structure functions are related by the
``5/18ths rule,''
$
F_2^{lN}/F_2^{\nu N}=5/18\{\; 1- 3/5[(s+\overline s)/(q+\overline q)]\; \}
$,
where the strange sea enters as a correction.
Comparison of structure function measurements from CCFR ($\nu$Fe) 
\cite{quintas,leung} with those
form SLAC ($e$D) \cite{emceffect}, NMC \cite{nmcmud} and BCDMS 
($\mu$D) \cite{bcdmsmud}, 
shows good agreement for $x>0.1$ but a small discrepancy 
is seen between the neutrino and muon results for lower $x$. The
source of this disagreement is under investigation, but may be due to the extra
axial vector component present in neutrino scattering \cite{donnachie}. 
In contrast, the recent global fits by the CTEQ
Collaboration (CTEQ1 distributions) \cite{cteq} have attributed the muon 
versus neutrino difference to an enhanced strange sea at low $x$.\footnote{The
more recent CTEQ2 distributions include a preliminary version of the NLO 
strange sea results presented here, but do not account for the structure
function discrepancy.}
This possibility is ruled
out by the measurements presented here.

\noindent{\bf Acknowledgement}

We thank M.A.G. Aivazis, F.I. Olness and W.-K. Tung for encouragement, useful
discussions and for providing computer code of their cross section calculation.
We thank the management and staff of Fermilab, and acknowledge the help of many
individuals at our home institutions. 
S.R.M. acknowledges the support of the Alfred P. Sloan and Cottrell 
Foundations.
This research was supported by the 
National Science Foundation and the Department of Energy
of the United States, who should be credited for their continued support of
basic research.

\bibliographystyle{unsrt}

\begin{thebibliography} {99}
\bibitem{distribs}
The parton distributions are parameterized at a reference scale $\mu_0^2$ by 
$x q_{NS}(x,\mu_0^2)  =   A_{NS} x^{\eta_1}(1-x)^{\eta_2} +
B_{NS}x^{\eta_3}$,
$x q_{SI}(x,\mu_0^2)  =   x q_{NS}(x,\mu_0^2) + A_{\rm sea}
(1-x)^{\eta_{\rm sea}}$ and
$x g(x,\mu^2_0)  =  A_g(1-x)^{\eta_g}$,  
and evolved to all other values of $\mu^2$ using
the GLAP \cite{glap}
evolution equations. 
$A_{NS}$ is constrained by the fermion conservation sum rule
$\int {dx\over x} \: xq_{NS}(x) = 3$,
and $A_g$ is constrained by the momentum sum rule
$\int dx \:\left[ xq_{SI}(x) + xg(x)\right] = 1$.
The final fit parameter is the strong coupling constant, given by
$\Lambda_{\overline{\rm MS}}$, for a total of eight fit parameters.
At $\mu_0^2=1\;{\rm GeV^2/c^2}$, the parameters are 
$\Lambda_{\overline{\rm MS}} = 0.210 \pm 0.021$ GeV/c,
$\eta_1=0.947 \pm 0.027$,
$\eta_2=3.81 \pm 0.059$,
$\eta_3= 0.405 \pm 0.047$,
$B_{NS}= 0.0210 \pm 0.0039$,
$A_{\rm sea}=1.404 \pm 0.047$,
$\eta_{\rm sea}=6.49 \pm 0.22$,
$\eta_g=3.18\pm 0.70$, where the errors are statistical only.
\end{thebibliography}
 
\end{document}